\documentclass[a4paper,11pt]{article}
\usepackage{jcappub}
\usepackage{aas_macros}
\usepackage{amsmath}

\makeatletter
\gdef\@fpheader{}
\makeatother

\newlength{\fullw}
\newlength{\halfw}
\newlength{\threefigw}
\newlength{\twofigw}
\newlength{\onefigw}
\newlength{\bigfigw}
\newlength{\roww}
\DeclareMathOperator{\sinc}{sinc}

\newcommand{\order}[1]{\mathcal{O}\!\left(#1\right)}

\newcommand{\boldmathsymbol}[1]{{\ensuremath{\boldsymbol{#1}}}}
\newcommand{\sss}[1]{{\scriptscriptstyle{#1}}}
\newcommand{\mean}[1]{\left\langle #1 \right\rangle}

\newcommand{\heaviside}[1]{\mathrm{\Theta}\!\left( #1 \right)}

\newcommand{\vect}[1]{\boldmathsymbol{#1}}
\newcommand{\negleft}{\negthinspace\left}
\newcommand{\dirac}[1]{\delta\negleft(#1\right)}

\newcommand{\Ho}{H_0}

\newcommand{\Mpc}{\mathrm{Mpc}}

\newcommand{\GeV}{\mathrm{GeV}}

\newcommand{\Hz}{\mathrm{Hz}}

\newcommand{\Mp}{M_\mathrm{Pl}}

\newcommand{\uTT}{\mathrm{\sss{TT}}}
\newcommand{\ud}{\mathrm{d}}
\newcommand{\uc}{\mathrm{c}}
\newcommand{\uh}{\mathrm{h}}

\newcommand{\us}{\mathrm{s}}

\newcommand{\ui}{\mathrm{i}}
\newcommand{\ue}{\mathrm{e}}
\newcommand{\ueq}{\mathrm{eq}}
\newcommand{\ugw}{\mathrm{gw}}

\newcommand{\umat}{\mathrm{mat}}
\newcommand{\urad}{\mathrm{rad}}
\newcommand{\umix}{\mathrm{mix}}
\newcommand{\ulcdm}{\mathrm{lcdm}}
\newcommand{\uini}{\mathrm{ini}}

\newcommand{\uppcl}{\mathrm{ppcl}}

\newcommand{\be}{\vect{e}}
\newcommand{\bepsilon}{\vect{\epsilon}}
\newcommand{\bx}{\vect{x}}
\newcommand{\by}{\vect{y}}
\newcommand{\bk}{\vect{k}}
\newcommand{\br}{\vect{r}}
\newcommand{\bkhat}{\be_k}
\newcommand{\bq}{\vect{q}}

\newcommand{\calF}{\mathcal{F}}

\newcommand{\calP}{\mathcal{P}}

\newcommand{\calH}{\mathcal{H}}
\newcommand{\calT}{\mathcal{T}}
\newcommand{\calU}{\mathcal{U}}

\newcommand{\calHo}{\calH_0}

\newcommand{\horizon}{d_\uh}
\newcommand{\corrini}{l_\uc}
\newcommand{\horizonini}{d_{\uh_\ui}}
\newcommand{\rhoinfty}{\rho_\infty}
\newcommand{\Nppcl}{N_{\uppcl}}

\newcommand{\x}{\kappa}
\newcommand{\xbar}{\bar{\x}}
\newcommand{\xini}{x_\uini}

\newcommand{\G}[1]{G_{#1}}
\newcommand{\Gmat}[1]{G^{\umat}_{#1}}
\newcommand{\Grad}[1]{G^{\urad}_{#1}}
\newcommand{\Gmix}[1]{G^{\umix}_{#1}}
\newcommand{\Glcdm}[1]{G^{\ulcdm}_{#1}}

\newcommand{\Ks}{K_{\us}}

\newcommand{\Ke}{K_{\ue}}

\newcommand{\Imu}{I_{\mu}}
\newcommand{\Idmu}{I_{\mu'}}
\newcommand{\Ix}{I_{\x}}
\newcommand{\Ixbar}{I_{\xbar}}

\newcommand{\OmegaR}{\Omega_{\urad}}
\newcommand{\OmegaM}{\Omega_{\umat}}

\newcommand{\OmegaGW}{\Omega_{\ugw}}

\newcommand{\calPh}{\calP_h}

\newcommand{\etaini}{\eta_\uini}
\newcommand{\etaeq}{\eta_\ueq}
\newcommand{\etanow}{\eta_{0}}
\newcommand{\etaenereq}{\bar{\eta}_{\ueq}}

\newcommand{\keq}{k_\ueq}

\newcommand{\xeq}{x_\ueq}

\newcommand{\zini}{z_\uini}

\newcommand{\U}{U}
\newcommand{\GU}{GU}

\begin{document}

\title{Stochastic gravitational waves from long cosmic strings}

\author[a]{Disrael Camargo Neves da Cunha,}
\author[a,b]{Christophe Ringeval}
\author[b]{and Fran\c{c}ois R. Bouchet}

\affiliation[a]{Cosmology, Universe and Relativity at Louvain (CURL),
  Institute of Mathematics and Physics, University of Louvain, 2 Chemin
  du Cyclotron, 1348 Louvain-la-Neuve, Belgium}

\affiliation[b]{Institut d'Astrophysique de Paris, 98bis boulevard Arago,
  75014 Paris, France}

\emailAdd{disrael.camargo@uclouvain.be}
\emailAdd{christophe.ringeval@uclouvain.be}
\emailAdd{bouchet@iap.fr}

\date{today}

\abstract{We compute the expected strain power spectrum and energy
  density parameter of the stochastic gravitational wave background
  (SGWB) created by a network of long cosmic strings evolving during
  the whole cosmic history. As opposed to other studies, the
  contribution of cosmic string loops is discarded and our result
  provides a robust lower bound of the expected signal that is
  applicable to most string models. Our approach uses Nambu-Goto
  numerical simulations, running during the radiation, transition and
  matter eras, in which we compute the two-point unequal-time
  anisotropic stress correlators. These ones act as source terms in
  the linearised equations of motion for the tensor modes, that we
  solve using an exact Green's function integrator.  Today, we find
  that the rescaled strain power spectrum $(k/\calHo)^2 \calPh$ peaks
  on Hubble scales and exhibits, at large wavenumbers, high frequency
  oscillations around a plateau of amplitude $100 (GU)^2$. Most of the
  high frequency power is generated by the long strings present in the
  matter era, the radiation era contribution being smaller.}


\maketitle

\section{Introduction}
\label{sec:intro}

The advent of gravitational wave astronomy has triggered a renewed
interest in the search for cosmic strings~\cite{Auclair:2019wcv,
  Gouttenoire:2019kij, Yonemaru:2020bmr,Jain:2020dct, Xing:2020ecz,
  Aurrekoetxea:2020tuw, Hernandez:2021vqh, Dunsky:2021tih,
  Gorghetto:2021fsn, Chun:2021brv}. These
objects can be of diverse cosmological origins, ranging from
topological defects formed during phase transitions in the Early
Universe~\cite{Kirzhnits:1972, Kobsarev:1974, Kibble:1976} to fundamental strings
living in higher warped multi-dimensional spaces~\cite{Witten:1985fp,
  Polchinski:2004yav, Sakellariadou:2009ev, Copeland:2009ga}.

Cosmic strings have been actively searched in various astronomical and
cosmological observables, for almost forty years, without triggering
any confirmed detection, see, e.g., Refs.~\cite{Gott:1984ef,
  Kaiser:1984iv, Vachaspati:1984gt, Hogan:1984is, Bouchet:1988,
  Accetta:1988bg, Bennett:1990ry, PhysRevD.45.1898, Hindmarsh:1994re,
  Damour:2000wa, Durrer:2001cg, Fraisse:2007nu, Ringeval:2010ca,
  Vachaspati:2015cma}. As of today, the strongest bound on the most
detectable strings comes from the stochastic gravitational wave
background limits set by the laser interferometers in the tens of
Hertz frequency window~\cite{KAGRA:2021kbb,
  LIGOScientific:2021nrg}. They report a maximal energy scale for
cosmic strings around $10^{12}\,\GeV$. Such an upper bound is
remarkably low, it deeply probes the energy scales that would be
associated with symmetry breaking occurring in the Grand Unified
Theories. At the same time, there is plenty of room for strings to be
formed at lower energies while a non-vanishing SGWB has been reported
in the nano-Hertz range~\cite{Arzoumanian:2020vkk,
  Antoniadis:2022pcn}. Moreover, the aforementioned limit only applies
to Nambu-Goto cosmic string loops having sizes given by the
Polchinski-Rocha scaling distribution~\cite{Ringeval:2005kr,
  Polchinski:2006ee, Rocha:2007ni, Lorenz:2010sm}, which, in addition,
are assumed to have a microstructure filled with ``kinks''. Kinks are
light-like moving shocks in the shape of the strings that repeatedly
collide and create spherical bursts of gravitational
waves~\cite{Ringeval:2017eww, Binetruy:2010cc}. The actual
microstucture of loops~\cite{Wachter:2016hgi, Wachter:2016rwc,
  Chernoff:2018evo} and the shape of their number density distribution
at small length scales are still a matter of
discussion~\cite{Ringeval:2005kr, Vanchurin:2005pa, Lorenz:2010sm,
  Blanco-Pillado:2013qja, Blanco-Pillado:2019tbi,
  Auclair:2019zoz}. This is particularly relevant as tiny changes in
these quantities have been shown to significantly impact the resulting
gravitational wave signal~\cite{Ringeval:2017eww,
  Auclair:2020oww}. Furthermore, the distribution of loops may be
model-dependent, and, at one extreme end, cosmic strings made of
Abelian Higgs field are observed to not produce stable loops at
all~\cite{Hindmarsh:2008dw, Hindmarsh:2021mnl}. In this situation,
only the long strings in a network would create persistent observable
signatures and are constrained by their effects onto the Cosmic
Microwave Background (CMB). The maximal possible energy scale
compatible with the Planck satellite data~\cite{ Ade:2013xla} is
around $10^{15}\,\GeV$~\cite{Ringeval:2010ca, Urrestilla:2011gr,
  Ringeval:2012tk, Lazanu:2014eya, 2017MNRAS.472.4081M, Ciuca:2017gca,
  Sadr:2017hfm}.  Let us mention that a non-scaling cosmic string
network, as one that would be formed during inflation, is even less
constrained~\cite{Yokoyama:1988zza, Jeong:2010ft, Kamada:2014qta,
  Ringeval:2015ywa, Guedes:2018afo, Cai:2021dgx}. Such a model dependence can
nevertheless be a virtue as more complex strings, e.g., the ones
endowed with currents, can potentially be detected via other means
than gravitational effects~\cite{Davis:1988ij, Brandenberger:1996zp,
  Cai:2012zd, Peter:2013jj, Hartmann:2017lno, 2017ApJ...850L..23M,
  Auclair:2020wse, Battye:2021dyq, Auclair:2021jud, Cyr:2022urs}.

In this paper, we focus on the stochastic gravitational wave
background produced by the long strings only, assuming that they are
the backbones of a scaling network. In order to consider most
realistic strings, we have performed new Nambu-Goto numerical
simulations, based on a modern version of the Bennett-Bouchet
code~\cite{Bennett:1989, Bennett:1990, Ringeval:2010ca,
  Ringeval:2012tk}, in the radiation and matter eras, as well as
during the transition from radiation to matter. While in scaling, a
Nambu-Goto cosmic string network exhibits universal statistical
properties and any observable quantity depends only on one
dimensionless parameter $\GU$, where $\U$ is the string energy
density, equals to the string tension, and $G$ the Newton
constant~\cite{Carter:2000wv}. The Nambu-Goto simulations allow us to
compute, without any approximation, the stress tensor associated with
the long strings, from which we derive the unequal-time two-point
correlators (UETC) of the anisotropic stress. These correlators act as
source terms in the linearised equations of motion for the spin-two
fluctuations propagating around a Friedmann-Lema\^itre-Robertson
Walker metric~\cite{Pen:1993nx, Durrer:1995sf, Magueijo:1995xj,
  Durrer:1997ep}. The method is inspired from the one presented in
Refs.~\cite{Figueroa:2012kw, Figueroa:2020lvo}, which has, up to now,
only been applied to global defects. However, as we discuss below, we
have paid special attention to keep all time and length scale
dependence in the computed waveforms in order to not wash-out any
oscillatory fine structure. This is particularly relevant in view of
the results derived in Ref.~\cite{daCunha:2021wyy}, which suggest that
cosmic strings would behave as ``singular'' sources at high
frequencies. In the following, we confirm these analytic expectations
while providing accurate numerical results for the overall shape of
the gravitational wave spectra. In particular, we find that the strain
power spectrum at large wavenumbers is driven by the long strings
evolving in the matter era, the contribution of the radiation-era
strings being a few orders of magnitude smaller.

The paper is organised as follows. In section~\ref{sec:method}, we
give some mathematical details on the Green's function method used to
compute the spectra from the UETC while discussing the accuracy of
assuming a pure radiation, or matter, era instead of the
$\Lambda$CDM solution. In section~\ref{sec:simus}, we give some
details on the Nambu-Goto numerical simulations and the method used to
compute the UETC while ensuring that they are close to the scaling
regime. Finally, in section~\ref{sec:spectra}, we present our main
results, which are the SGWB spectra of the rescaled strain $k^2/(12
\calHo^2) \calPh(k)$, and the energy density parameter $\OmegaGW(k)$,
generated by long cosmic strings. Finally, we conclude in
section~\ref{sec:conclusion}.

\section{Gravitational waves of cosmological origin}
\label{sec:method}
In this section, we briefly recap the linearised equations for the
tensor modes created and propagated in a
Friedmann-Lema\^{\i}tre-Robertson-Walker (FLRW) metric. We then
discuss the computation of the associated Green's function in the
$\Lambda$CDM model.

\subsection{Linearised equations of motion}

Assuming a spatially flat metric of the form
\begin{equation}
\ud s^2 = a^2(\eta)\left\{-\ud \eta^2 + \left[\delta_{ij} +
  h_{ij}(\eta,\bx) \right] \ud x^i \ud x^j  \right\},
\label{eq:pertmetric}
\end{equation}
where $h_{ij}(\eta,\bx)$ is a gauge-invariant divergenceless and
traceless tensor (Roman indices running only on spatial dimensions),
the Einstein equations, linearised at first order in $h_{ij}$, read
\begin{equation}
h_{ij}'' + 2 \dfrac{a'}{a} h_{ij}' - \Delta h_{ij} = \dfrac{2}{\Mp^2} a^2 \varPi_{ij}.
\label{eq:einstein}
\end{equation}
Here a prime denotes differentiation with respect to $\eta$, the
conformal time, the Laplacian operator is over the comoving spatial
coordinates $\bx$ and $\Mp$ stands for the reduced Planck mass. The
source term $\varPi_{ij} = \delta T_{ij}^{\uTT}(\eta,\bx)/a^2$ is the
traceless and transverse anisotropic part of the linearised stress
tensor $\delta T_{\mu\nu}$ at the origin of the
perturbations~\cite{Mukhanov:1990me}. In these equations, all scalars
and vectors have been assumed to vanish as we are only focused in the
generation and propagation of gravitational waves. In order to solve
equation~\eqref{eq:einstein} in the presence of long cosmic strings,
we need to compute $\varPi_{ij}(\eta,\bx)$ and invert the differential
operator to get $h_{ij}(\eta,\bx)$. Moreover, we are interested in the
statistical properties of the stochastic background generated by the
superimposition of all these waves and one has to construct the
two-point correlation functions of these quantities.

In the helicity basis, and Fourier space, the strain can be decomposed
as~\cite{Zaldarriaga:1996xe, Alexander:2004wk}
\begin{equation}
h_{ij}(\eta,\bx)= \dfrac{1}{(2 \pi)^3} \int_{-\infty}^{\infty}
\sum_{r=-2,+2} h_{r}(\eta,\bk) \epsilon_{ij}^r(\bkhat)
   e^{\imath \bk
    \bx} \ud^3\bk,
   \label{eq:FTh}
\end{equation}
where the $\epsilon_{ij}^r(\bkhat)$ are the component of the helicity
basis tensor. In the Fourier spherical coordinate system
$(\bkhat,\be_1,\be_2)$, they read
\begin{equation}
\bepsilon^{\pm 2} = \dfrac{1}{2} \left(\be_1 \pm \imath \be_2 \right)
\otimes \left(\be_1 \pm \imath \be_2 \right),
\end{equation}
and verifies
\begin{equation}
\epsilon_{ij}^{r *} \epsilon^{ij}_s = \delta^r_s.
\end{equation}
Moreover, one has the relation $\epsilon_{ij}^{\pm 2}(-\bkhat) =
\epsilon_{ij}^{\pm 2*}(\bkhat)$ to ensure that $h_{ij}(\eta,\bx)$ are
real quantities. In the following, we will work with the mode function
\begin{equation}
\mu_r \equiv a(\eta)h_r,
\label{eq:mudef}
\end{equation}
which, from equation~\eqref{eq:einstein}, is solution of
\begin{equation}
\mu_r''(\eta,\bk) + \left(k^2 - \dfrac{a''}{a} \right) \mu_r(\eta,\bk)
= \dfrac{2}{\Mp^2} a^3 \varPi_r(\eta,\bk).
\label{eq:muevol}
\end{equation}
This equation can be exactly solved using the retarded Green's function,
$\G{\xi}(\eta,\bk)$, solution of the same equation with a source term
distribution in $\delta(\eta-\xi)$. Assuming $\G{\xi}(\eta,\bk)$ to be
known, and the source $\varPi_r$ to vanish at times $\eta < \etaini$,
the mode function, and its time derivative, are then given by
\begin{equation}
\begin{aligned}
  \mu_r(\eta,\bk) & = \dfrac{2}{\Mp^2 k} \int_{\etaini}^{\eta} k
  \G{\xi}(\eta,k) a^3(\xi) \varPi_r(\xi,\bk) \ud \xi,\\
  \mu_r'(\eta,\bk) & = \dfrac{2}{\Mp^2} \int_{\etaini}^{\eta}
\G{\xi}'(\eta,k) a^3(\xi) \varPi_r(\xi,\bk) \ud \xi.
\label{eq:musol}
\end{aligned}
\end{equation}
where $\G{\xi}'$ stands for $\partial \G{\xi}(\eta,k)/\partial
\eta$. Notice that these equations encode both the generation of
gravitational waves, in the domains for which $\varPi_r(\xi,\bk) \ne 0$,
and the propagation to the observer thanks to the Green's
functions. We now turn to the determination of these functions.

\subsection{Computation of the Green's functions}

\label{sec:green}

There are known analytical solutions for the retarded Green's
functions associated with equation~\eqref{eq:muevol} and we simply
quote the results. If the scale factor behaves as in a pure radiation
era $a(\eta) \propto \eta$, one has~\cite{Caprini:2018mtu}
\begin{equation}
k \Grad{\xi}(\eta,k) = \sin\left[k(\eta - \xi)\right] \heaviside{\eta-\xi}.
\label{eq:greenrad}
\end{equation}
For a matter era that instantaneously succeeds the radiation era one
has $a(\eta) \propto (\eta + \etaeq)^2$ and~\cite{daCunha:2021wyy}
\begin{equation}
\begin{aligned}
k \Gmat{\xi}(\eta,k) &= \left\{ \sin\left[k(\eta - \xi)\right] -
\dfrac{k(\eta-\xi)}{k(\eta+\etaeq) \, k(\xi+\etaeq)}
\cos\left[k(\eta-\xi)\right] \right. \\ & \left. +
\dfrac{1}{k(\eta+\etaeq) \, k(\xi+\etaeq)} \sin\left[k(\eta-\xi)\right] \right\} \heaviside{\eta-\xi}.
\end{aligned}  
\label{eq:greenmat}
\end{equation}
Let us stress that, in this last expression, $\etaeq$ is not a free
parameter. Denoting by $\calHo=\calH(\etanow)$ the conformal Hubble parameter today,
it is defined as~\cite{daCunha:2021wyy}
\begin{equation}
  \etaeq \equiv \dfrac{\sqrt{\OmegaR}}{\calHo \OmegaM}\,,
  \label{eq:etaeqdef}
\end{equation}
to ensure continuity of the scale factor, and its derivative, at the
instantaneous transition between radiation and matter. In particular,
its numerical value differs (by a few) from the conformal time at
which there would be the actual equality between the energy density of
matter and radiation.

\begin{figure}
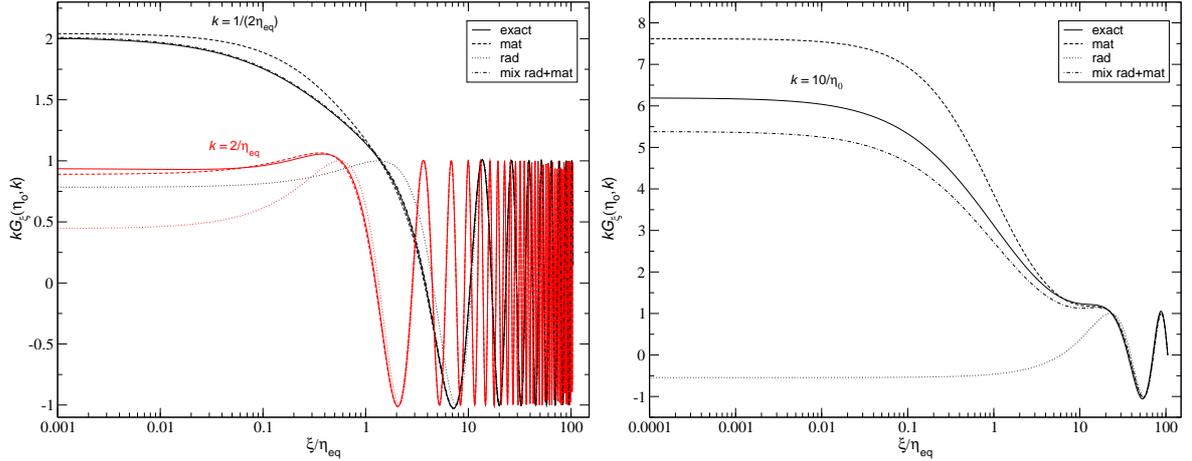

\begin{center}
  \includegraphics[width=\twofigw]{green_exact_vs_matrad_etaeq}
  \includegraphics[width=\twofigw]{green_exact_vs_matrad_etanow}
  \caption{Green's functions $k\G{\xi}(\eta=\etanow,k)$ associated
    with the evolution of tensor modes, up to now. They are plotted
    for various values of the wavenumber $k$, as a function of the
    time parameter $\xi$, expressed in unit of $\etaeq$. The
    right-most value of $\xi/\etaeq \simeq 105$ corresponds to today
    with $\xi=\etanow$. The label ``exact'' stands for the Green's
    function computed for the $\Lambda$CDM model while ``rad+mat'' for
    a scale factor driven by a mixture of radiation and matter. The
    labels ``rad'' and ``mat'' refer to the analytical solutions
    obtained by assuming a pure radiation era, and, a matter era
    patched to a radiation era, given by equations~\eqref{eq:greenrad}
    and \eqref{eq:greenmat}, respectively. The most commonly used
    approximation in the literature (``rad'') \emph{always} produces
    erroneous values at early times. It remains acceptable only on the
    time domains for which the mode remains deeply sub-Hubbles (see
    also figure~\ref{fig:greeninrad}). On the contrary, the matter era
    solution is surprisingly good, even during the radiation era.}
\label{fig:greentoday}
\end{center}
\end{figure}

\begin{figure}
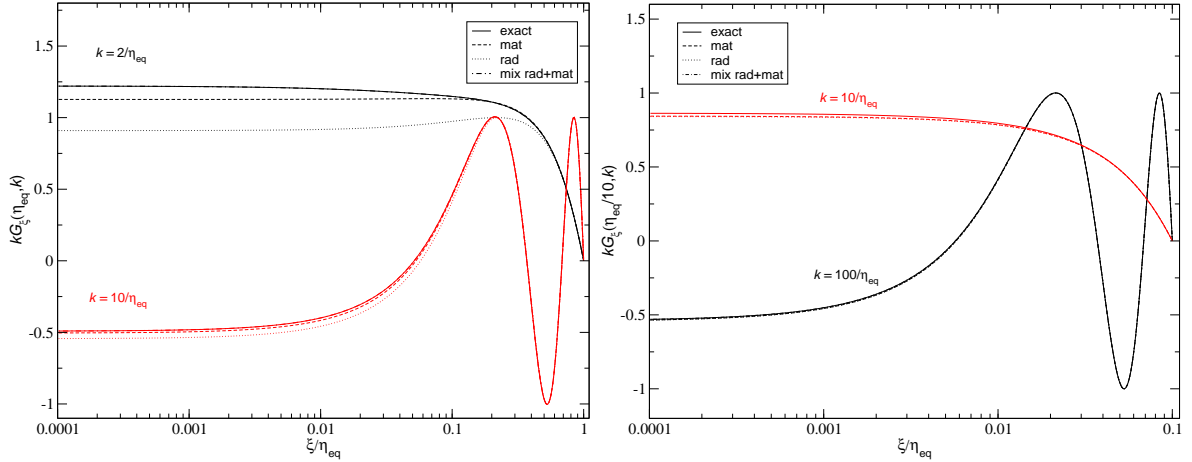

\begin{center}
  \includegraphics[width=\twofigw]{greenintrans}
  \includegraphics[width=\twofigw]{greeninrad}
  \caption{The Green's function $k\G{\xi}(\etaeq,k)$ (left panel) and
    $k\G{\xi}(\etaeq/10,k)$ (right panel), evaluated for a time
    parameter $\xi$ exploring the radiation era. As one may expect,
    the Green's function computed for a mixture of radiation and
    matter is undistinguishable from the exact one. However, for
    $\eta$ set at the transition (left plot), the matter era Green's
    function is still a better approximation than the one derived
    assuming pure radiation. Deeper into the radiation era, at
    $\eta=\etaeq/10$ (right plot), all approximations are
    undistinguishable from the exact solution.}
\label{fig:greeninrad}
\end{center}
\end{figure}

The evolution of the scale factor $a(\eta)$ in presence of a
mixture of radiation and matter can be exactly solved from the
Friedmann-Lema\^{\i}tre equations and reads
\begin{equation}
a(\eta) = \dfrac{\OmegaM \calHo^2}{4} \eta^2 + \sqrt{\OmegaR} \calHo \eta.
\label{eq:amix}
\end{equation}
Plugging this expression into equation~\eqref{eq:muevol}, the associated Green's
function $\Gmix{\xi}(\eta,k)$ is solution of
\begin{equation}
  \dfrac{\ud^2 \left[k\Gmix{y}(x)\right]}{\ud x^2} + \left(1 - \dfrac{2}{x^2 + 4
    \xeq x} \right) k\Gmix{y}(x) = \delta(x-y),
\label{eq:greenmix}
\end{equation}
where we have made the change of variables
\begin{equation}
x \equiv k \eta, \qquad y = k \xi, \qquad \xeq =  k\etaeq\,.
\end{equation}
In spite of a simple form, equation~\eqref{eq:greenmix} has no
analytic solutions and must be solved numerically. Using the Wronskian
method, for each value of $x$ and $y$, one needs two numerical
integrations, with different initial conditions, to allow for the
determination of the Green's function. These solutions are represented
in figures~\ref{fig:greentoday} and \ref{fig:greeninrad}, where they
are compared to the approximations given by
equations~\eqref{eq:greenrad} and \eqref{eq:greenmat} as well as to
the $\Lambda$CDM solution (see below).

We have also numerically determined the Green's function $k
\Glcdm{\xi}(\eta,k)$, that we refer to as ``exact'' in these
figures. It is obtained by considering the scale factor evolution
associated with the $\Lambda$CDM model, the cosmological parameter
values having being fixed to the ones favoured by the Planck satellite
measurements~\cite{Planck:2018vyg}. In this situation, there is no
analytical solution for the scale factor evolution $a(\eta)$. The
determination of $k \Glcdm{\xi}(\eta,k)$, at fixed $\xi$, $\eta$ and
$k$, therefore requires, in parallel to the two aforementioned
numerical integrations, the computation of $a''(\eta)/a(\eta)$
appearing in equation~\eqref{eq:muevol}. This one is obtained by
numerically integrating the Friedmann-Lema\^{\i}tre equations.

In figure~\ref{fig:greentoday}, we have represented the Green's
functions $k\G{\xi}(\eta,k)$, evaluated at the present time
$\eta=\etanow$, for various values of $k$, as a function of the time
parameter $\xi$. As can be seen in equation~\eqref{eq:musol}, once
multiplied by the anisotropic stress, this is the quantity of interest
to perform the convolution integral. The left panel shows two modes
entering the Hubble radius around equality, $k=1/(2 \etaeq)$ (black
curves) and $k=2/\etaeq$ (red curves). For each of them, we have
plotted the numerical $\Lambda$CDM solution (labelled ``exact''), the
numerical solution of equation~\eqref{eq:greenmix} (``rad+mat''),
equation~\eqref{eq:greenrad} (``rad'') and
equation~\eqref{eq:greenmat} (``mat''). In most of the matter era
($\xi/\etaeq \gg 1$), the modes are sub-Hubble and freely propagate,
all Green's functions are the same. However, at earlier times, when
the modes were of wavelength close, or greater, than the Hubble radius
at that time (for $\xi/\etaeq \lesssim 1$), deviations appear. The
differences between $\Glcdm{\xi}$ and $\Gmix{\xi}$ are barely visible
whereas the pure radiation era Green's function $\Grad{\xi}$ is the
most inaccurate. For the mode $k=1/(2\etaeq)$, differences between
$\Grad{\xi}$ and the others are noticeable for $\xi/\etaeq \lesssim
10$. Interestingly, $\Gmat{\xi}$ remains a relatively good
approximation of the exact solution even for the super-Hubble modes in
the radiation era. The right panel of figure~\ref{fig:greentoday}
shows the Green's functions for a mode of wavelength close to the
Hubble radius today, $k=10/\etanow$. Differences between the exact
solution $\Glcdm{\xi}$ and the mixture radiation and matter
$\Gmix{\xi}$ are now visible. This is expected as the cosmological
constant domination is very recent in the history of the Universe and
can only show up at time $\eta \simeq \etanow$. The radiation era
Green's function still dramatically fails while the matter era one
overestimates the expected values, up to $20\%$ in the radiation era.

It is important to notice that these conclusions are
$\eta$-dependent. Therefore, in figure~\ref{fig:greeninrad} we have
represented the very same functions, $k\G{\xi}(\eta,k)$, but evaluated
at times which are either during the transition $\eta=\etaeq$ (left
panel), or during the radiation era at $\etaeq/10$ (right panel). The
differences are less pronounced than for $\eta=\etanow$, but, still,
for all modes becoming super-Hubble close to $\xi=\etaeq$,
$\Grad{\xi}$ differs from the others. Confined within the radiation
era (right panel), all functions lead to the same values, including
the matter era one.

In conclusion, using $\Gmix{\xi}$ (mixture radiation and matter) to
describe the generation and propagation of gravitational waves is
accurate at all times and for all modes which are sub-Hubble today,
namely $k > 10/\etanow$. Next in accuracy is, quite surprisingly, the
fully analytic matter era Green's function $\Gmat{\xi}$ given by
equation~\eqref{eq:greenmat}. It remains within a ten percent error
margin from the exact solution, and mostly deviates for Hubble-like
modes today, as expected from the present acceleration of the Universe
which is not accounted for. The worse of all is $\Grad{\xi}$. It is
accurate only for modes created, propagated and measured within the
radiation era.

Let us finally notice that the accuracy of the final result, namely
the value of $\mu_r(\eta,\bk)$, also depends on the support of the
anisotropic stress $\varPi_r(\xi,\bk)$. If this one vanishes in the
domains in which the Green's function is inaccurate, the final result
may very well be valid, but this is no trivial a priori and needs to
be checked.

\subsection{Propagating gravitational wave power spectra}

\label{sec:propsgw}

The superimposition of all gravitational waves generated by a network
of long cosmic strings produces a stochastic background, which is, at
leading order, isotropic. Comparison with observation requires to
determine its statistical properties, and, in this paper, we focus on
the two-point correlation functions. In Fourier space, using the
statistical invariance by translation, the unequal time correlator of
the strain is of the form
\begin{equation}
\mean{h_r^*(\eta_1,\bk) h_r(\eta_2,\bq)} \equiv \dfrac{(2\pi)^3}{V}
\dirac{\bk-\bq} P_{h_r}(\eta_1,\eta_2,\bk),
\label{eq:htwopointdef}
\end{equation}
where we have made the (infinite) volume factor $V$ explicit. Statistical
isotropy implies that
$P_{h_r}(\eta_1,\eta_2,\bk)=P_{h_r}(\eta_1,\eta_2,k)$ and the strain
correlation function in real space can be expressed as
\begin{equation}
\mean{h_{ij}(\eta_1,\bx) h^{ij}(\eta_2,\bx+\by)} = \int_0^\infty \dfrac{\ud q}{q} 
\calP_h(\eta_1,\eta_2,q) \sinc(q y),
\label{eq:straincorr}
\end{equation}
where the (spherical) strain power spectrum we are interested in reads
\begin{equation}
\calP_{h}(\eta_1,\eta_2,k) = \sum_r \calP_{h_r}(\eta_1,\eta_2,k)
\equiv \dfrac{k^3 V}{2\pi^2} \sum_r P_{h_r}(\eta_1,\eta_2,k).
\label{eq:calPdef}
\end{equation}
Similarly, the spatial two-point correlators of the time derivative
$h_r'$ is expressed, in Fourier space, as the dimensionless parameter
\begin{equation}
\OmegaGW(\eta_1,\eta_2,k) \equiv \dfrac{\sum_r
  \calP_{h_r'}(\eta_1,\eta_2,k)}{12 \calH(\eta_1) \calH(\eta_2)}\,,
\label{eq:OmegaGW}
\end{equation}
where $\calP_{h_r'}$ is defined from $P_{h_r'}$ in the exact same
manner as in equations~\eqref{eq:htwopointdef} and \eqref{eq:calPdef}.

From equations~\eqref{eq:mudef}, both $P_{h_r}$ and $P_{h_r'}$ can be
derived from the power spectra of $\mu_r$ and $\mu_r'$. One
has~\cite{daCunha:2021wyy}
\begin{equation}
\begin{aligned}
  P_{h_r}(\eta_1,\eta_2,k) & =
  \dfrac{P_{\mu_r}(\eta_1,\eta_2,k)}{a(\eta_1) a(\eta_2)}\,, \\
  P_{h_r'} (\eta_1,\eta_2,k) & = H(\eta_1) H(\eta_2) \left[
    \dfrac{P_{\mu_r'}(\eta_1,\eta_2,k)}{\calH(\eta_1) \calH(\eta_2)} + P_{\mu_r}(\eta_1,\eta_2,k) -
    \dfrac{P_{\x_r}(\eta_1,\eta_2,k)}{\calH(\eta_1)} \right. \\ &
    \left. - 
    \dfrac{P_{\xbar_r}(\eta_1,\eta_2,k)}{\calH(\eta_2)} \right].
\end{aligned}
\label{eq:Phs}
\end{equation}
where we have defined
\begin{equation}
\mean{{\mu_r'}^*(\eta_1,\bk) \mu_r(\eta_2,\bq)} =  \dfrac{(2\pi)^3}{V}
\dirac{\bk-\bq} P_{\x_r}(\eta_1,\eta_2,\bk),
\end{equation}
and $P_{\xbar_r}(\eta_1,\eta_2,\bk) =
P_{\x_r}^*(\eta_2,\eta_1,\bk)$. From equation~\eqref{eq:musol}, one has
\begin{equation}
\begin{aligned}
  P_{\mu_r}(\eta_1,\eta_2,\bk) & = \dfrac{4}{k^2 \Mp^4}
  \int_{\etaini}^{\eta_1} \ud \xi \int_{\etaini}^{\eta_2} \ud \xi' k
  \G{\xi}^*(\eta_1,k) \, k \G{\xi'}(\eta_2,k) a^3(\xi) a^3(\xi')
  \mean{\varPi_r^*(\xi,\bk) \varPi_r(\xi',\bk)},
  \\ P_{\mu_r'}(\eta_1,\eta_2,\bk) & = \dfrac{4}{\Mp^4}
  \int_{\etaini}^{\eta_1} \ud \xi \int_{\etaini}^{\eta_2} \ud \xi'
      {\G{\xi}'}^*(\eta_1,k) \G{\xi'}'(\eta_2,k) a^3(\xi) a^3(\xi')
      \mean{\varPi_r^*(\xi,\bk) \varPi_r(\xi',\bk)},
      \\ P_{\x_r}(\eta_1,\eta_2,\bk) & = \dfrac{4}{k \Mp^4}
      \int_{\etaini}^{\eta_1} \ud \xi \int_{\etaini}^{\eta_2} \ud \xi'
          {\G{\xi}'}^*(\eta_1,k) \, k \G{\xi'}(\eta_2,k) a^3(\xi)
          a^3(\xi') \mean{\varPi_r^*(\xi,\bk) \varPi_r(\xi',\bk)}.
\end{aligned}
\label{eq:Pmus}
\end{equation}
As a result, the source terms needed to uniquely determine all the
power spectra are the unequal time anisotropic stress
correlators. Using translational invariance and statistical isotropy,
they can be redefined as
\begin{equation}
\mean{\varPi_r^*(\xi,\bk) \varPi_r(\xi',\bq)} = \dfrac{(2 \pi)^3}{V}
\dirac{\bk - \bq}
\dfrac{\U^2 \calT_r(\xi,\xi',k)}{a^2(\xi) \sqrt{\xi} \, a^2(\xi') \sqrt{\xi'}}\,,
\label{eq:piredef}
\end{equation}
Here, we have explicitly factorised the typical energy density scale
$\U^2$ and $\calT_r(\xi,\xi',k)$ is a dimensionless
function. The interest of expressing the anisotropic stress
correlators as in equation~\eqref{eq:piredef} is that, when the source
terms are in the so-called scaling regime, one has~\cite{Durrer:1997ep, Wu:1998mr, Bevis:2006mj,
  Lazanu:2014xxa}
\begin{equation}
  \calT_r(\xi,\xi',k) = \calU_r(k\xi,k\xi').  
\label{eq:scalingdef}
\end{equation}
A network of cosmic strings reaches scaling in both the radiation and
matter era, with, however, different functions, say,
$\calU^\urad_r(x,x')$ and $\calU^\umat_r(x,x')$, where $x=k\eta$. Let
us notice that equation~\eqref{eq:piredef} is valid even if the
anisotropic stress does not assume a scaling form and we will be using
$\calT_r(\xi,\xi',k)$ during the transition between the radiation and
the matter eras.

Plugging equation~\eqref{eq:piredef} into equation~\eqref{eq:Pmus}, one
can simplify equation~\eqref{eq:calPdef} into
\begin{equation}
\calPh(\eta_1,\eta_2,k) = 128 (\GU)^2 \Imu(x_1,x_2,k),
\label{eq:calPhx}
\end{equation}
where the integral
\begin{equation}
\Imu(x_1,x_2,k) \equiv \dfrac{1}{a(\eta_1)a(\eta_2)} \int_{\xini}^{x_1}
  \ud x \int_{\xini}^{x_2} \ud x' \Ks(x_1,x) \Ks(x_2,x')
  \dfrac{a\left(\frac{x}{k}\right) a\left(\frac{x'}{k}\right)}{\sqrt{x
      x'}}  \calT\left(\frac{x}{k},\frac{x'}{k},k\right).
\label{eq:Imu}
\end{equation}
In this expression, we have defined the dimensionless parameters $x_1
= k\eta_1$, $x_2 = k\eta_2$ and $\xini = k \etaini$ as well as $\calT
= \sum_r \calT_r$. The integration kernel for the strain, $\Ks$, is
the rescaled Green's function, see equation~\eqref{eq:greenmix}, and
reads
\begin{equation}
\Ks(x_i,x) \equiv k \G{x}(x_i).
\label{eq:ksdef}
\end{equation}
One can also express $\OmegaGW$ in terms of similar integrals. From
equations~\eqref{eq:OmegaGW} and \eqref{eq:Pmus}, one gets
\begin{equation}
\begin{aligned}
  \OmegaGW(\eta_1,\eta_2,k) & = \dfrac{32}{3} \left(\GU
  \right)^2\left[\dfrac{k^2}{\calH(\eta_1) \calH(\eta_2)}
    \Idmu(x_1,x_2,k) + \Imu(x_1,x_2,k) \right. \\ & \left.
    -\dfrac{k}{\calH(\eta_1)} \Ix(x_1,x_2,k) -
    \dfrac{k}{\calH(\eta_2)} \Ixbar(x_1,x_2,k) \right],
\end{aligned}
\label{eq:omegax}
\end{equation}
with
\begin{equation}
\begin{aligned}
\Idmu(x_1,x_2,k) & \equiv \dfrac{1}{a(\eta_1)a(\eta_2)} \int_{\xini}^{x_1}
  \ud x \int_{\xini}^{x_2} \ud x' \Ke(x_1,x) \Ke(x_2,x')
  \dfrac{a\left(\frac{x}{k}\right) a\left(\frac{x'}{k}\right)}{\sqrt{x
      x'}}  \calT\left(\frac{x}{k},\frac{x'}{k},k\right),\\
  \Ix(x_1,x_2,k) & \equiv \dfrac{1}{a(\eta_1)a(\eta_2)} \int_{\xini}^{x_1}
  \ud x \int_{\xini}^{x_2} \ud x' \Ke(x_1,x) \Ks(x_2,x')
  \dfrac{a\left(\frac{x}{k}\right) a\left(\frac{x'}{k}\right)}{\sqrt{x
      x'}}  \calT\left(\frac{x}{k},\frac{x'}{k},k\right),
\end{aligned}
\label{eq:Idmus}
\end{equation}
and $\Ixbar(x_1,x_2,k) = \Ix^*(x_2,x_1,k)$. In these two integrals,
the energy kernel, $\Ke$, is defined as
\begin{equation}
\Ke(x_i,x) \equiv \left.\dfrac{\partial \left[k
    \G{x}(y)\right]}{\partial y}\right|_{y=x_i} = \dfrac{\partial
  \Ks(x_i,x)}{\partial x_i}\,.
\end{equation}
As a result, we have the following relations
\begin{equation}
\Idmu(x_1,x_2,k) = \dfrac{\partial^2\Imu(x_1,x_2,k)}{\partial x_1
  \partial x_2}\,,\qquad \Ix(x_1,x_2,k) = \dfrac{\partial\Imu(x_1,x_2,k)}{\partial x_1}\,,
\label{eq:partialx1x2}
\end{equation}
and, a priori, only $\Imu(x_1,x_2,k)$ needs to be determined for
computing both $\calPh$ and $\OmegaGW$. However, in the absence of an
analytical expression for $\Imu$, it is easier and more accurate to
compute the three integrals $\Imu$, $\Idmu$ and $\Ix$ separately. The
integration kernels $\Ks$ and $\Ke$ being the Green's function of
$\mu$ and $\mu'$, their determination has already been discussed in
section~\ref{sec:green}. In the next section, we present the method
used to numerically estimate the function $\calT$.

\section{Anisotropic stress correlators}

\label{sec:simus}

\subsection{Nambu-Goto numerical simulations}

\begin{figure}
\begin{center}
  \includegraphics[width=\bigfigw]{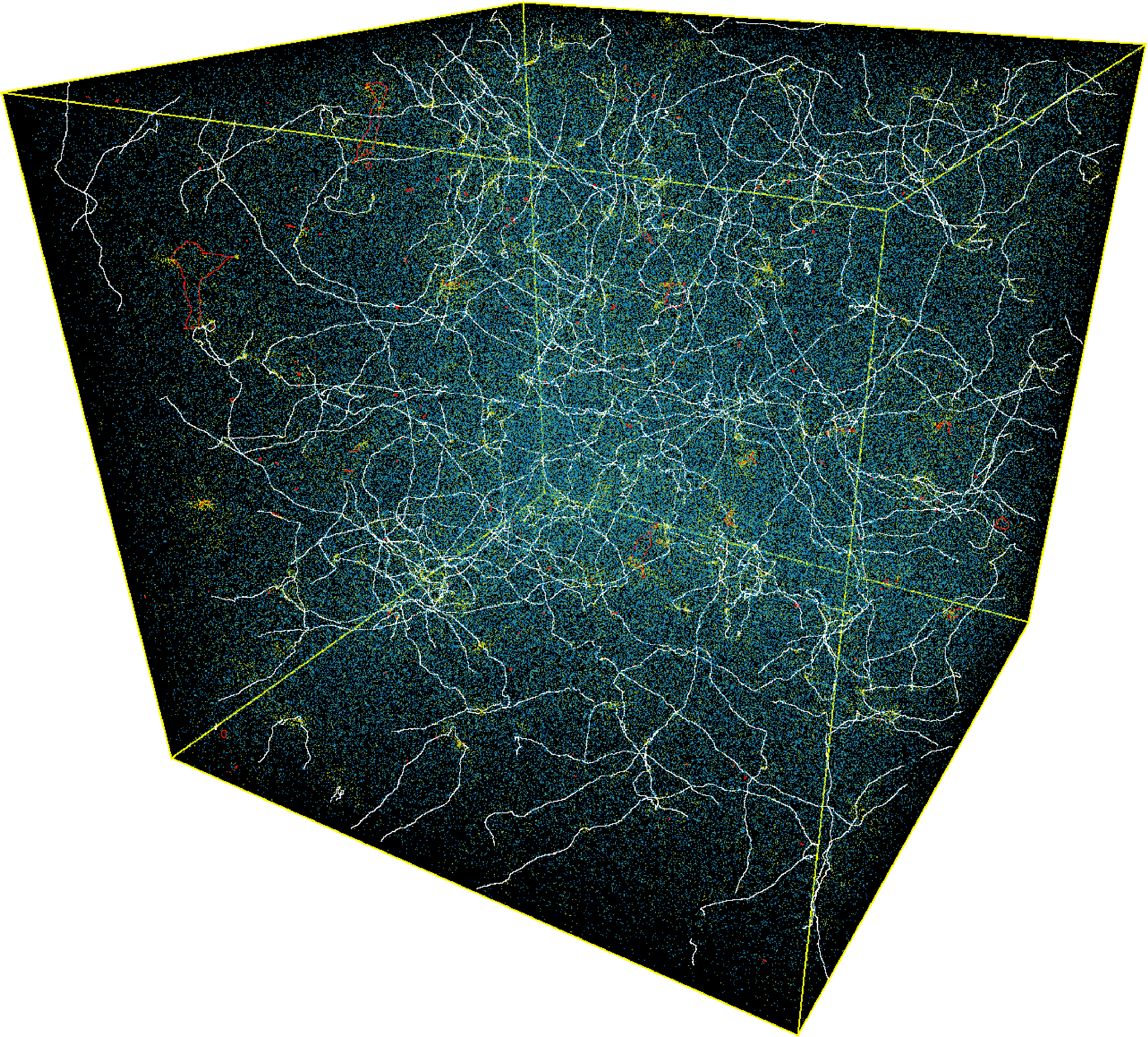}
  \caption{Snapshot of the cosmic string network during the transition
    era when the conformal horizon occupies $60\%$ of the fixed
    comoving volume $(100\corrini)^3$. For this work, the stress
    tensor includes only the contribution of the long strings, defined
    as being longer than the horizon size. They are represented in
    white in the picture. Loops in scaling are represented in red
    whereas freshly formed loops coming from the fragmentation of
    larger structures are represented in yellow. The blue ``fog'' is
    made of tiny older loops. There are about two million strings in
    this picture, they are all numerically evolved till the end of the
    run.}
\label{fig:snapshot}
\end{center}
\end{figure}

\begin{figure}
\begin{center}
  \includegraphics[width=\onefigw]{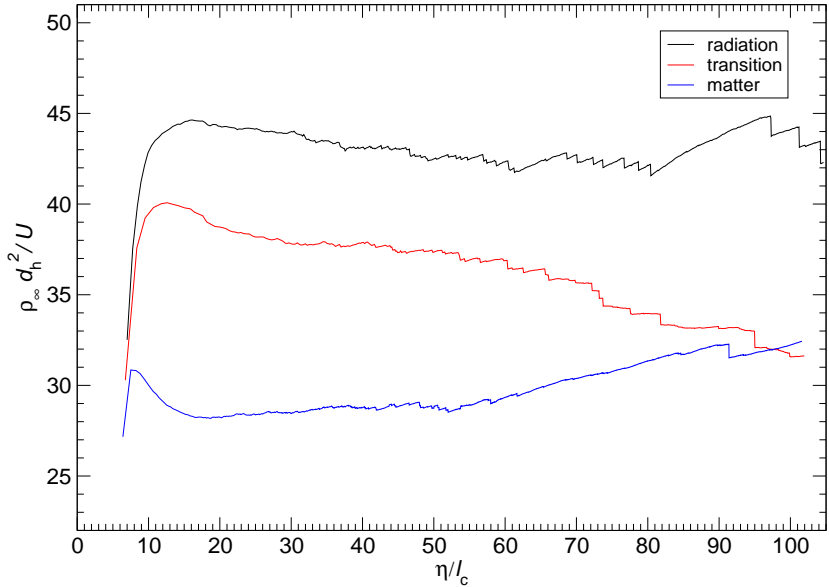}
  \caption{Rescaled energy density of the long strings, defined as
    being longer than the horizon size, as a function of the
    simulation conformal time $\eta$ (in unit of the initial
    correlation length $\corrini$). The curves correspond to the three
    runs made in the radiation, transition and matter era. After a
    fast growing phase for $\eta \simeq \horizonini$ (leftmost side),
    the long strings reach a scaling regime during which $\rhoinfty
    \horizon^2$ remains stationary. Notice however the slow drift of
    the transition era that interpolates between the larger energy
    density attractor of the radiation era towards the one of the
    matter era. The ``saw teeth'', and slight upward drift, visible at
    large $\eta$ are due to a cosmic variance effect. Towards the end
    of the run, the total number of super-horizon strings in the
    simulation decreases to only a few, and, each collision, or
    reconnection event, induces a significant shift, and uncertainty,
    in the energy density estimator.}
\label{fig:enerlong}
\end{center}
\end{figure}

In order to numerically evaluate the unequal time anisotropic stress
correlators of equation~\eqref{eq:piredef}, we have run new
numerical simulations of Nambu-Goto strings in FLRW spacetimes, based
on a modern version of the Bennett-Bouchet code. See
Refs.~\cite{Bennett:1989, Bennett:1990, Ringeval:2010ca,
  Ringeval:2012tk} for a detailed description of the code.

The simulations evolve a network of Nambu-Goto cosmic strings, created
from Va\-chas\-pati-Vilenkin (VV) initial
conditions~\cite{Vachaspati:1984}, in a fixed unity comoving volume
with periodic boundary conditions. The simulations can only be run
during a finite amount of time before the non-trivial topology starts
to be felt and they are stopped when the horizon fills the whole
volume. Normalising the scale factor to unity at the beginning of the
run, there are three adjustable physical parameters that fix the
initial state of the network. The first is the VV correlation length,
which has been set to $\corrini = 10^{-2}$. In other words, the
comoving box has a size of $100\corrini$.  A typical VV realisation
with $\corrini = 10^{-2}$ contains about $20000$ strings. The second
parameter is the initial size of the horizon $\horizonini$, which, in
our units, is also the initial conformal time. This parameter
determines the initial number density of long strings, defined as
being of length greater than the horizon size. Finally, the last one
is the maximal amplitude of a transverse random velocity field and has
been set to $0.1c$ so as to optimise relaxation towards scaling. In
addition to the initial conditions, another adjustable physical
parameter is $\etaenereq$, the conformal time of equality between the
energy density of matter and radiation.  In addition to these physical
parameters, the code has various numerical parameters that control
discretization accuracy. Among them, let us stress that these
simulations evolve strings which are initially discretized with
$\Nppcl=20$ points per correlation length. In other words, the initial
box has a numerical resolution of $2000^3$ points, which goes on
increasing thanks to adaptive mesh refinement
methods~\cite{Bennett:1989, Bennett:1990}. By the end of the runs,
each simulation contains about two million strings, mostly under the
form of loops, whose motion, collisions and fragmentations are all
accounted for, at all times.

We have run three simulations, one in the radiation era with
$\horizonini=0.070$ ($\etaenereq \to \infty$), one over the transition
radiation to matter, with $\horizonini=0.0675$ and $\etaenereq=0.110$
(see figure~\ref{fig:snapshot}) and one in the matter era with
$\horizonini=0.064$ ($\etaenereq \to 0$). These numbers are set such
that the relaxation time of the network towards scaling is small and,
for $\etaenereq$, to put a slightly larger dynamical range of the
simulation onto the transition to matter era. In
figure~\ref{fig:enerlong}, we have represented the energy density of
super-horizon strings $\rhoinfty \horizon^2/U$ as a function of the
conformal time. Long string scaling is reached when this quantity
remains stationary. As can be seen in these plots, there is a
transient period at the beginning of the runs (leftmost side) during
which the energy density rapidly grows. Then, it slows downs and
relaxes towards a stationary value for the radiation and matter
era. For the transition era, after the initial fast growth, another
long relaxation takes place during which the energy density slowly
drifts from radiation-like values towards the matter era attractor.

\subsection{Stress tensor computation}

The cosmic string code has been extended to compute the Fourier
transform of the Nambu-Goto stress tensor of any object present within
the simulation volume.

For this purpose, we have defined a fixed three-dimensional comoving
grid, here with $1024^3$ points, over which we estimate the ten
independent components of the Nambu-Goto stress tensor in the
transverse and temporal gauge (see, e.g., Ref.~\cite{Ringeval:2010ca}
for details). The numerical estimator uses a cloud-in-cell (CIC)
method~\cite{Sefusatti:2015aex} where the additive contribution from
each little piece of string is distributed over its eight nearest grid
neighbours, with a distance weighting factor. This is done at various
time steps during the run while a filter allows us to include only the
long string contribution. Let us stress that the loops are kept within
the Nambu-Goto simulation, so as to ensure that their backreaction
over the long strings is properly accounted for. They are just
discarded in the stress tensor computation. Once the stress tensor
over the grid is estimated, we perform a three-dimensional Fourier
transform using the ``Fastest Fourier Transform of the West'' (FFTW)
library~\cite{FFTW05}. This one is then deconvolved with the CIC
window function in order to recover a unsmoothed estimator of the
stress tensor. Finally, we extract $T_{ij}^\uTT(\eta,\bk)$, the
traceless and divergenceless part of the stress tensor and project it
onto the helicity basis of equation~\eqref{eq:FTh}. The outcome of
this process is a set of three-dimensional Fourier transformed
anisotropic stresses
\begin{equation}
\varPi_r(\eta_p,\bk_q) = \epsilon^{r*}_{i j} \dfrac{\delta
  T_\uTT^{ij}(\eta_p,\bk_q)}{a^2(\eta_p)},
\label{eq:varpinum}
\end{equation}
for both helicities $r=\pm 2$, at discretely sampled times $\eta_p$,
and discrete wavenumbers $k_q = \pm 2\pi q$, $q
\in\{0,\dots,512\}$. These data are finally dumped as files (in the
``fits'' format) and are used later on for constructing the unequal
time correlators (see next section). In order to maintain a reasonable
storage space for these files, at around $1$ Terabytes per simulation,
the time sampling has been chosen to get about $40$ time steps
$\eta_p$ by the end of each run.

\subsection{Relaxation to scaling}

\label{sec:relax}

\begin{figure}
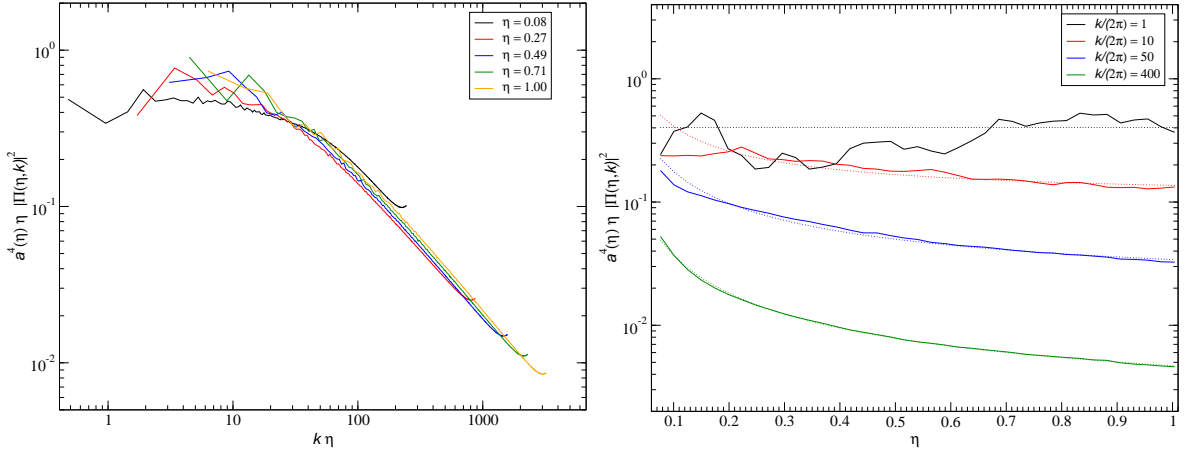

\begin{center}
  \includegraphics[width=\twofigw]{uetc_pibarpi}
  \includegraphics[width=\twofigw]{relax_pibarpi}
  \caption{Rescaled equal-time correlator $a^4 \eta |\varPi|^2$, in the
    matter era, as a function of $k\eta$ (left panel), evaluated at
    different conformal times $\eta$. In scaling $a^4 \eta |\varPi|^2$
    is stationary and the curves on the left panel tend to
    superimpose. The excess of power close to the Nyquist frequency is
    an aliasing artefact (it is removed in the final calculation). The
    right panel shows the same equal-time correlators but plotted as a
    function of $\eta$, for different wavenumbers $k$. Although the
    large length scale modes (small values of $k$) are in scaling at
    early times in the simulation, the smallest length scales are
    still relaxing by the end of the run. We use a fit to these
    relaxation curves to discard, at any given time, modes too far
    from scaling (see text).}
\label{fig:pibarpi}
\end{center}
\end{figure}

From the discretely sampled anisotropic stresses
$\varPi_r(\eta_p,\bk_q)$, we construct an estimator of the unequal
time correlator by using the statistical isotropy in Fourier space,
i.e., we define
\begin{equation}
 \mean{\varPi_r^*(\eta_1,k) \varPi_r(\eta_2,k)} =
 \dfrac{1}{4\pi}\int_0^{2\pi} \ud \phi \int_0^{\pi} \ud \theta \sin
 \theta \,
 \varPi_r^*(\eta_1,\bk) \varPi_r(\eta_2,\bk),
\label{eq:uetcavg}
\end{equation}
where $(\theta,\phi)$ stand for the spherical angles in Fourier
space. This procedure has the advantage to require only one simulation
to determine the correlators. However it is associated with a
scale-dependent variance, the small values of $k$ suffering from
larger uncertainties due to the low number of modes $\bk_q$ close to
the origin (see figure~\ref{fig:pibarpi}).

Among all the sampled values of $\eta_p$ during the simulation, we
would like to quantify how much the correlators computed from
equation~\eqref{eq:uetcavg} are close to scaling and not too much
affected by the initial conditions. From equations~\eqref{eq:piredef} and
\eqref{eq:scalingdef}, in scaling, the rescaled equal-time correlator
$a^4(\eta) \eta |\varPi(\eta,k)|^2$  should remain stationary while
depending only on $k\eta$.

In the left panel of figure~\ref{fig:pibarpi}, we have plotted this
quantity as a function of $x=k \eta$ and at different times $\eta_p$
during the simulation. The spectrum is flat on large scales and turns
to a power law decay at large $x$, with a power-law exponent slowly
approaching $-1$ towards the end of the run. This shape matches the
one expected for a distribution of one-dimensional objects, the change from a
plateau-like shape to a power-law decay precisely occurring at the
correlation length of the string network~\cite{Wu:1998mr}.

Ignoring the expected cosmic variance effects at small $x$, the
different curves are quasi-superimposed one onto the others, with,
however, a slowly evolving tail at large wavenumbers. Comparing the
power-law tail of the anisotropic stress plotted for $\eta=0.71$ to
the one at the end of the run ($\eta=1$) suggests that scaling is
indeed reached up to $x \lesssim 200$.  Above this value, the
amplitude of the tail seems to be slightly changing with time
suggesting that these small scale modes are still relaxing. This is
more visible in the right panel of figure~\ref{fig:pibarpi}, in
which we have plotted $a^4(\eta) \eta |\varPi(\eta,k)|^2$ as a
function of $\eta$, and for different modes $k$ (solid
curves). Although the large scale modes (small values of $k$) are in
scaling from very early times, high wavenumbers such as $k=800\pi$ are
indeed relaxing all over the simulation time.

In the same figure, the dotted curves represent, for each mode $k$,
the best fit obtained by assuming the relaxation function to be a
power-law decay towards a constant value. These fits allow us to
determine, at any given time during the simulation, by how much a
given mode of the anisotropic stress correlator differs from its
expected asymptotic value. Calling the relative difference
$\delta_k(\eta)$, we can now use it as a criterion to discard, or not,
the mode when constructing the full unequal-time correlator of
equation~\eqref{eq:piredef} (see next section). Clearly, discarding
the values $\delta_k(\eta)< \delta$ with $\delta$ very small would
essentially remove all modes at all times and some compromise has to
be done between scaling accuracy and limiting the noise. Unless
otherwise stated, we allow each mode to be, at most, from
$\delta=50\%$ to the fitted and extrapolated asymptotic value. This
criterion also gives the typical error we will have on the final
result due to the slow convergence towards scaling of the smallest
length scales.

\subsection{Machine-learned correlators}
\label{sec:rbf}

From the previous discussion, we have at our disposal the correlators
$\mean{\varPi_r^*(\xi_l,k_q)\varPi_r(\xi_p,k_q)}$ for some discrete values
of $\xi_l$, $\xi_p$ and $k_q$. From equation~\eqref{eq:piredef}, these numbers
give us a sampling of the function $\calT(\xi,\xi',k) = \sum_r\calT_r(\xi,\xi',k)$.

\subsubsection{Correlators in scaling}

\label{sec:Urbf}

\begin{figure}
\begin{center}
  \includegraphics[width=\bigfigw]{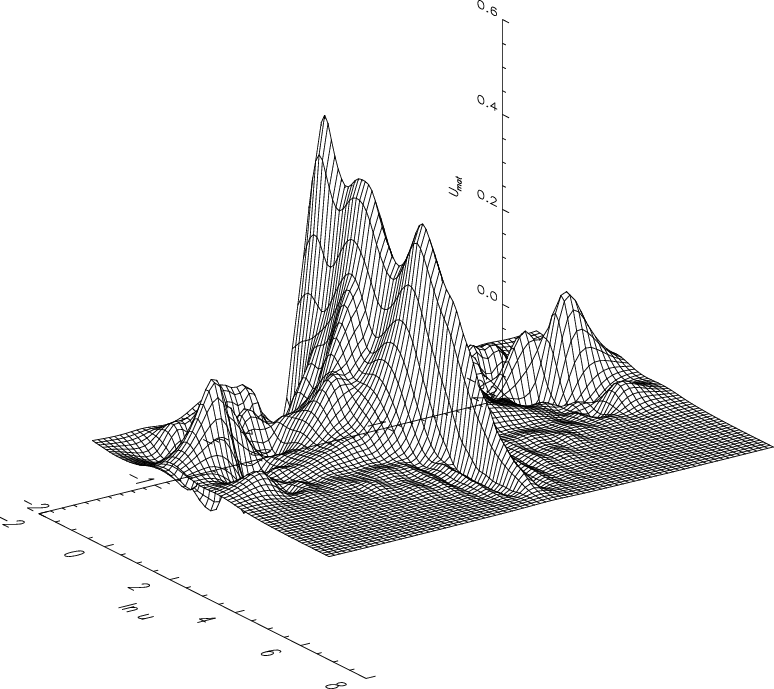}
  \caption{The scaling function $\calU^{\urad}(\ln u,\ln v)$
    reconstructed using a radial basis function decomposition based on
    the sparse data computed in the Nambu-Goto numerical
    simulations. The unequal-time correlator is maximal along the $u$
    axis, i.e., at equal times. On Hubble length scales ($\ln u <0$),
    time correlations persist over, typically, one Hubble time. Notice
    the damped off-diagonal oscillations which are expected from
    causality. On small scales (large $\ln u$), the equal-time
    correlator decays as $k^{-1}$.}
\label{fig:Urad}
\end{center}
\end{figure}

\begin{figure}
\begin{center}
  \includegraphics[width=\bigfigw]{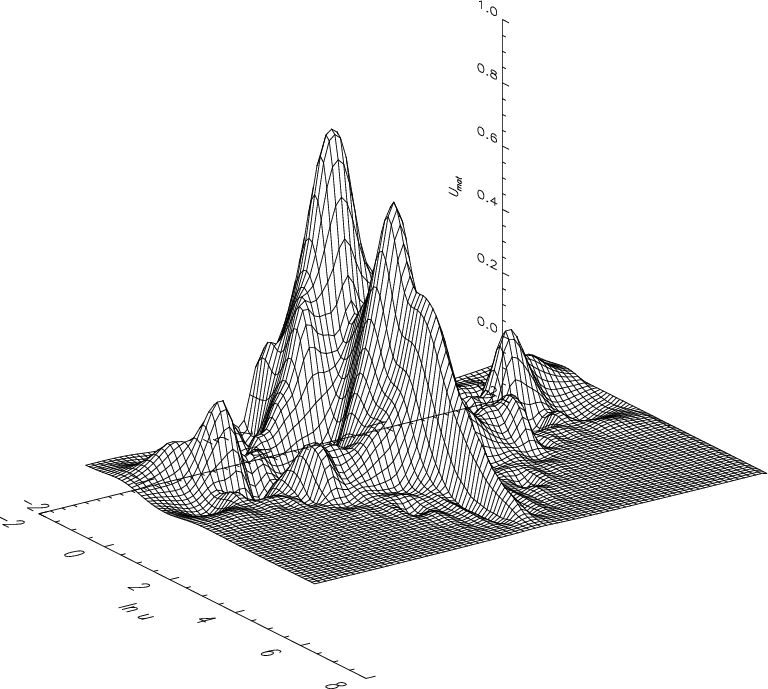}
  \caption{The scaling function in the matter era, $\calU^{\umat}(\ln
    u,\ln v)$, reconstructed using a radial basis function
    decomposition. Notice the higher amplitude than the radiation-era
    function plotted in figure~\ref{fig:Urad}.}
\label{fig:Umat}
\end{center}
\end{figure}
Let us first focus on the radiation and matter eras. The scaling
ensures that $\calT(\xi,\xi',k)=\calU(x,x')$, where, as before,
$x=k\eta$. One expects this function to be maximal along the diagonal
$x=x'$ as the correlations should decay at large unequal times. It is
therefore convenient to work with the new variables
\begin{equation}
u \equiv \sqrt{x x'} = k \sqrt{\eta \eta'}, \qquad v \equiv
  \sqrt{\dfrac{x'}{x}} = \sqrt{\dfrac{\eta'}{\eta}},
\label{eq:uvdef}
\end{equation}
where the wavenumber dependence is now only in $u$\footnote{Not to be
confused with the $(u,v)$ variables appearing in
Ref.~\cite{Durrer:2001cg}.}. Moreover, since $u$ and $v$ are positive
and may assume variations over a few orders of magnitude, we will be
working in the $(\ln u, \ln v)$ space. In order to numerically
determine the function $\calU$ from irregular gridded data, we have
used some simple machine-learning method based on a radial basis
function (RBF) decomposition~\cite{Broomhead:1988}. Basically, this
consists in decomposing the function $\calU(\ln u,\ln v)$ as a
weighted sum of isotropic functions, here inverse quadratic, centered
over a set of nodes in the $(\ln u,\ln v)$ space. More specifically,
denoting the vector $\br\equiv(\ln u,\ln v)$, we fit the correlator by
the function
\begin{equation}
\calF(\br) = \sum_{i=1}^N \varpi_i \left[1 +
  \dfrac{\left(\br-\br_i\right)^2}{\lambda^2} \right]^{-1},
\label{eq:rbffit}
\end{equation}
where $\lambda$ is a scale parameter and $N$ the total number of nodes
used. The quantities $\br_i$ are the location of each node in this
two-dimensional space. They have been evenly distributed along a
two-dimensional grid encompassing the whole data range. The weights
$\varpi_i$ are then computed as to minimize the difference between the
$\calF(\br_j)$ and $\calU(\br_j)$, where the $\br_j$ are running on
the actual data coming from the cosmic string simulation. For the
$\br_j$, we could have used the raw sparse data, but, in order to
minimize the noise, best results were obtained by binning the raw
data (using a nearest neighbourg method) before feeding them to a singular
value decomposition solver to invert Eq.~\eqref{eq:rbffit} and
determining the weights $\varpi_i$. Finally, the scale parameter
$\lambda$ is determined by minimising the difference between the raw
data (unbinned) and the radial basis function predictor. Notice that,
at each fitting iteration of $\lambda$, one needs to recompute all the
weights. As a matter of fact, such a decomposition is strictly
equivalent to a single layer neural network having as many neurons as
RBF nodes. Less than percent accuracy in the decomposition is
typically achieved with a few hundred nodes ($N\gtrsim 400$).

As an illustration, figure~\ref{fig:Urad} shows the resulting function
$\calU^{\urad}(\ln u,\ln v)$ over the range of values probed by the
Nambu-Goto simulation. Most of the power is at equal times, $x=x'$,
along the $\ln v =0$ direction. A broad peak is visible on Hubble
scales ($\ln u \simeq 0$) while the ``crest'' of the correlator starts
decaying as $k^{-1}$ at large $\ln u$, the transition between the two
behaviour occurring at the length scales associated with the mean
inter-string distance. Non-vanishing off-diagonal values ($\ln v \ne
0$) are essentially located on Hubble scales, and, typically last for
one Hubble time. Let us remark the presence of some oscillations in
these regions. They are expected by causality, indeed, if correlations
abruptly disappear outside the light cone in real space, then
oscillations should be present in Fourier
space~\cite{Turok:1996ud}. The matter era scaling function
$\calU^{\umat}(\ln u,\ln v)$ has been represented in
figure~\ref{fig:Umat} and is of similar shape. Let us however remark
the higher overall amplitude and the larger off-diagonal structures at
large scales.

Let us notice that the very same correlator has been derived for
Abelian Higgs string in Ref.~\cite{Daverio:2015nva} as to compute
their induced CMB anisotropies. In that reference, the authors fit the
off-diagonal part by a slowly decaying function, and, such a slow
decay is similar to the off-diagonal behaviour we observe but on
sub-Hubble scales only. However, on Hubble-scales, the presence of the
off-diagonal oscillations prevented us to use this method and this is
one of the reasons for having used the radial basis function
method. It is nevertheless interesting to notice that Figure~8 of
Ref.~\cite{Daverio:2015nva} shows some off-diagonal structures as
well, on Hubble scales, which could very well be oscillatory but
cut-off by the smaller dynamical range of the Abelian Higgs field
simulations compared to our Nambu-Goto simulations.

Although the range of values in $u$ and $v$ accessible in the
Nambu-Goto simulation covers a few orders of magnitude (see
figures~\ref{fig:Urad} and \ref{fig:Umat}), the integrals of
equation~\eqref{eq:Imu} and \eqref{eq:Idmus} are over a much larger
domain. The interest of having used a RBF decomposition is that the
functions $\calU(\ln u, \ln v)$ can be evaluated outside the fitted
domain. The asymptotic behaviour is there driven by the one of the
radial basis functions and it decays to zero as a sum of inverse
quadratic functions. However, as shown in Ref.~\cite{daCunha:2021wyy},
the precise asymptotic form of the equal-time correlator at large $\ln
u$ (along the $\ln v = 0$ direction) determines the high wavenumber
shape of the final gravitational wave power spectra. As such, for the
large values of $\ln u$ outside the fitting domain, we have
implemented a power law extrapolation of the correlators $\calU(\ln
u,\ln v = 0) \propto u^{-1}$ while keeping the RBF decomposition along
the $\ln v$ direction\footnote{Under some conditions, the precise
decay of the off-diagonal correlations does not significantly alter
the large wavenumber limit~\cite{daCunha:2021wyy}.}. On the largest
scales, i.e., for large negative values of $\ln u$, we have
implemented a constant extrapolation as one can show that the
correlator should be constant in this limit~\cite{Durrer:1997ep}.

\subsubsection{Transition era}

\label{sec:Trbf}

As can be seen in figure~\ref{fig:enerlong}, the energy density of
long strings during the transition era slowly drifts from
radiation-like values towards the one expected in a pure matter
era. One should expect a similar behaviour for the anisotropic stress
correlator and this motivates us to rewrite the function
\begin{equation}
\calT(\eta,\eta',k) = \calT^\umix(\ln u,\ln v,w),
\end{equation}
where we have introduced the new dimensionless parameter
\begin{equation}
w \equiv \dfrac{\eta+\eta'}{2 \etaenereq}.  
\end{equation}
From the previous section, one expects
\begin{equation}
  \calT^\umix(\ln u,\ln v,w \to 0) = \calU^{\urad}(\ln u,\ln
  v), \qquad 
  \calT^\umix(\ln u,\ln v,w \gg 1) = \calU^{\umat}(\ln u,\ln v),
\end{equation}
such that $w$ encodes the relaxation time from the radiation-era
correlator to the matter-era one. In practice, the matching to radiation-
and matter-era is made at a finite value of $w$, one which is accessible within the
transition-era simulation, here at $w=0.8$ and $w=8$.

Numerically, the sparse data extracted from the transition-era
simulation have been again fitted with a RBF decomposition, but now,
in the three dimensions $(\ln u,\ln v, w)$. Convergence requires a few
thousand nodes, as opposed to a few hundred for the correlators in
scaling, such that the computations are slightly more demanding (see
section~\ref{sec:Urbf}). But the main problem is to determine which
domains in the space $(\eta,\eta',k)$ are not too much affected by the
initial conditions. Because there is no scaling any more, one cannot
use the stationarity of $a^4(\eta) \eta |\varPi(\eta,k)|^2$ as a
criterion of being non-contaminated by the initial
conditions. Noticing that the Nambu-Goto simulation starts with a
scale factor evolution which is close to the one of the radiation era,
we have chosen to take cuts in the $k$ and $\eta$ domains identical to
the ones of a pure radiation era, with the same value $\delta = 50\%$
(see section~\ref{sec:relax}). As we show in the next section, the
contribution of the transition era to the final result remains however
subdominant.

\section{Gravitational waves from long strings}
\label{sec:spectra}

\begin{figure}
\begin{center}
  \includegraphics[width=\onefigw]{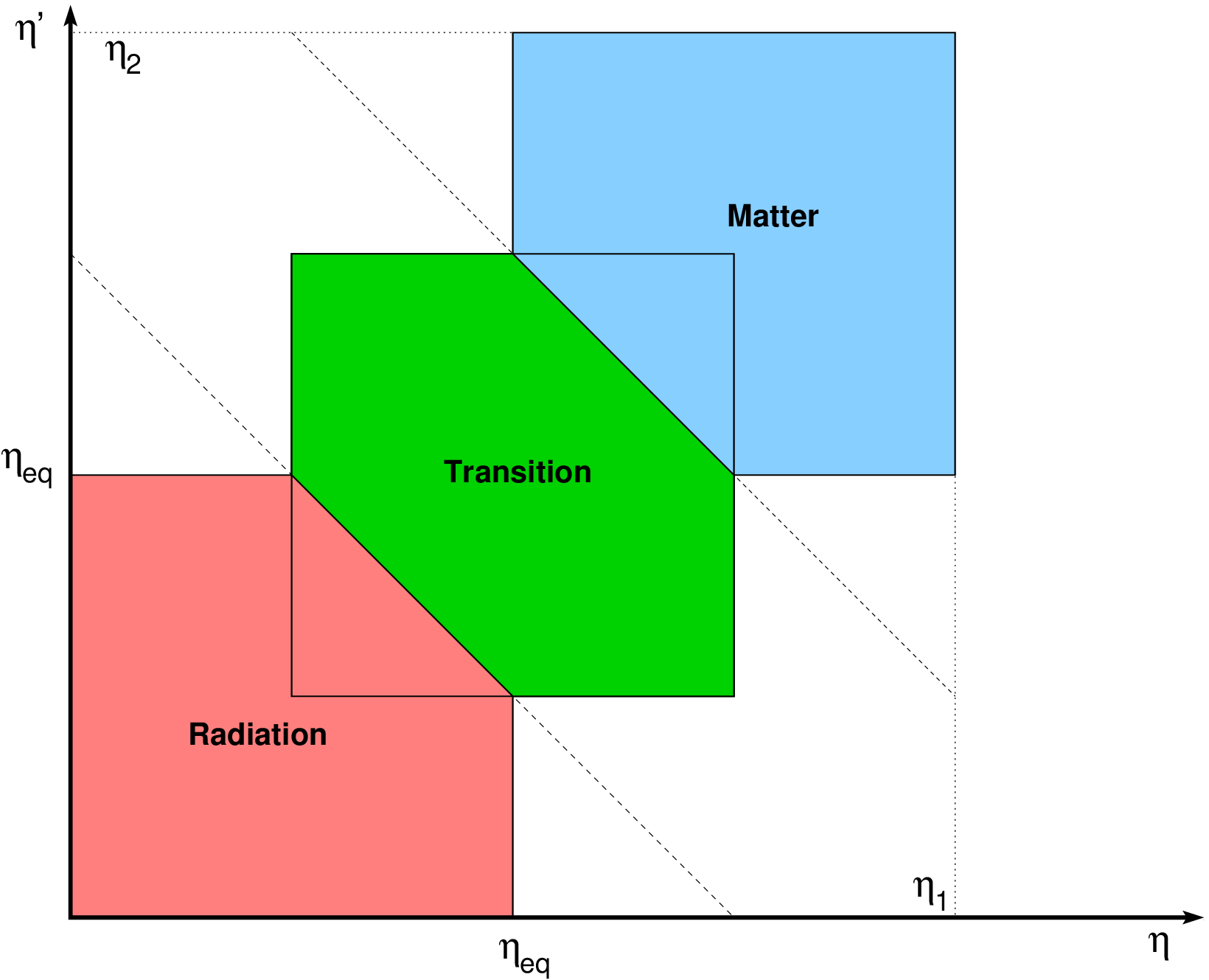}
  \caption{Conformal time domains (coloured squares) in which the
    anisotropic stress unequal-time correlators are determined from
    the radiation-era, transition-era and matter-era Nambu-Goto
    numerical simulations. As can be seen in figures~\ref{fig:Urad}
    and \ref{fig:Umat}, off-diagonal correlations ($\eta \ne \eta'$)
    rapidly decay such that the coloured squares are covering all of
    the physically relevant regions. However, there are several
    manners to pave the three numerical simulations. Here, we have
    chosen to patch them at fixed value of the parameter
    $w=(\eta+\eta')/(2\etaenereq)$ (dashed lines).}
\label{fig:domains}
\end{center}
\end{figure}

\subsection{Computing resources}

The strain power spectrum $\calPh(\eta_1,\eta_2,k)$, and the energy
density parameter $\OmegaGW(\eta_1,\eta_2,k)$, evaluated at some
possibly unequal-times, $\eta_1$ and $\eta_2$, require the computation
of the kernels $\Ks(x_i,x)$, $\Ke(x_i,x)$ and of the function
$\calT(x/k,x'/k,k)$ at all past times $\etaini \le \eta \le
\eta_{1,2}$. These quantities indeed appear in the three integrals
$\Imu$, $\Idmu$ and $\Ix$ given in equations~\eqref{eq:Imu} and
\eqref{eq:Idmus}. They are two-dimensional integrals over the domain
$[\xini,x_1]\times[\xini,x_2]$ that need to be evaluated for each
values of the wavenumber $k$. Figure~\ref{fig:domains} sketches the
integration domain and its covering by the Nambu-Goto numerical
simulations. In practice, we have fixed the value of $\xini=k\etaini$
by assuming that the string network has been formed in the early
Universe at a redshift $\zini=10^{30}$, which corresponds to $\etaini
\simeq 10^{-28}/\calHo$. As shown in section~\ref{sec:green}, for the
exact $\Lambda$CDM model, the scale factor $a(x/k)$ appearing in the
integrand can only be obtained by numerically integrating the
Friedmann-Lema\^{\i}tre equations. The convolution kernels, evaluated
at a given $(x_i,x)$, require two additional numerical integrations of
the differential operator~\eqref{eq:muevol}. As such, computing these
integrals is numerically demanding. For this purpose, we have used the
$\texttt{CUBA}$ library~\cite{Hahn:2004fe}, which is a Monte-Carlo
type integrator, i.e., points in the space $(x,x')$ are randomly
picked to evaluate the integrals. For each mode $k$, a target accuracy
of $10^{-3}$ may typically demand drawing $5\times 10^{9}$ points,
especially for large values of $k$ where the integrand fastly
oscillates and this requires as much numerical integration of the
scale factor and of the convolution kernels.

In order to speed-up the calculations, we have parallelised the code
using the Message Passing Interface ($\texttt{MPI}$) to allow for the
computation of each wavenumber $k$ to be performed on a different
machine. Moreover, we have used another level of parallelism using the
$\texttt{OpenMP}$ directives to distribute the integrand evaluation
over the different processors available on a single machine. A final
speed-up of two has been reached by using the $\texttt{AVX2}$ vector
registers for the most inner operation. Finally, in view of the
results of section~\ref{sec:green}, for all wavenumbers $k \ge 10
\calHo$, the calculations are simplified by neglecting the
cosmological constant effects and using the analytic
form~\eqref{eq:amix} of the scale factor. This allows us to skip the
numerical integration of the Friedmann-Lema\^{\i}tre equations, but,
still, for these modes, the Green's functions $\G{x}^{\umix}(x_i)$ can
only be obtained by numerically integrating
equation~\eqref{eq:greenmix}.

\subsection{Results}

\begin{figure}
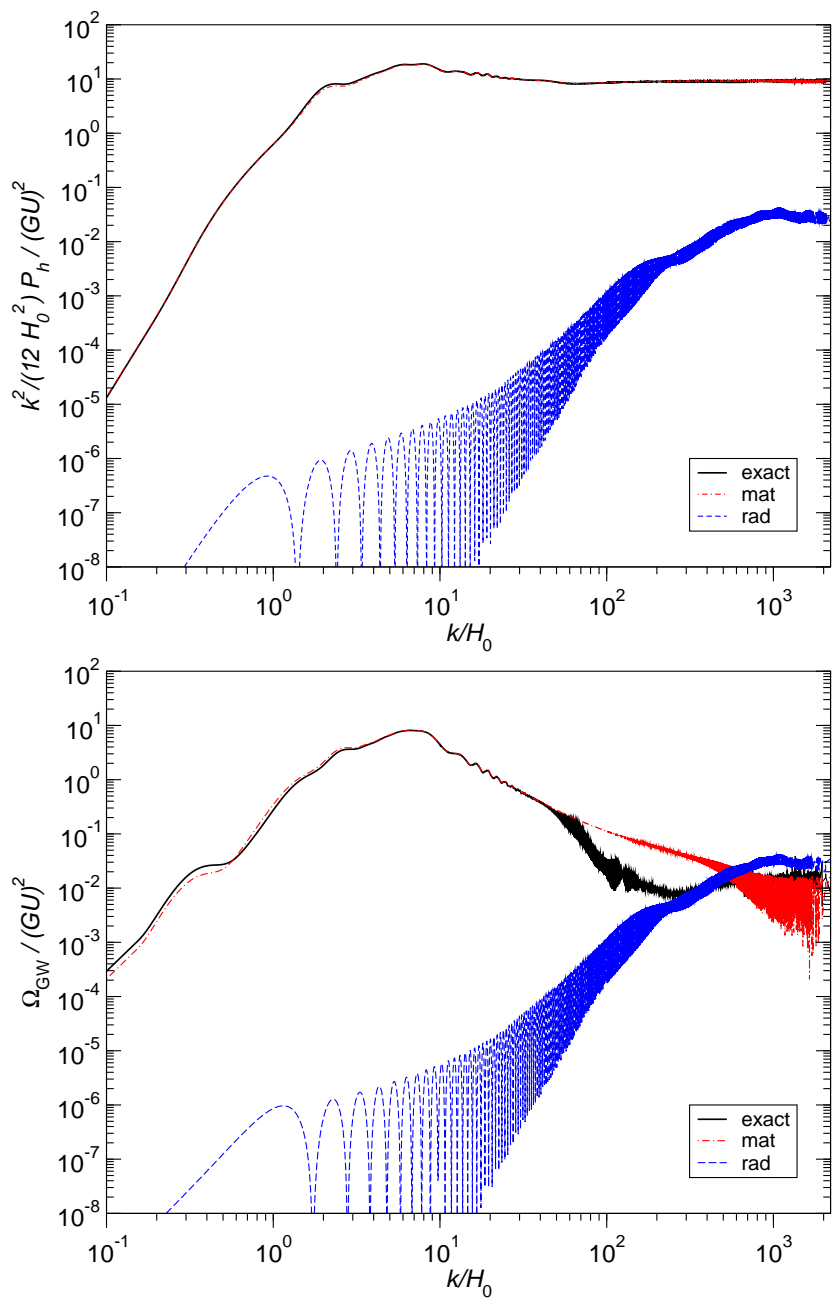

\begin{center}
  \includegraphics[width=\onefigw]{k2calPh}
  \includegraphics[width=\onefigw]{OmegaGW}
  \caption{The GW rescaled strain power spectrum $k^2/(12
    \calHo^2)\calPh(\etanow,\etanow,k)$ (top panel) and the energy
    density parameter $\OmegaGW(\etanow,\etanow,k)$ (bottom panel),
    today, generated by long cosmic strings. In both panels, the black
    solid curve is the exact numerical result using the $\Lambda$CDM
    scale factor evolution and $\Lambda$CDM Green's function for the
    tensor modes. The dashed curves show some semi-analytical
    approximations of the matter era (red) and radiation era (blue)
    contributions assuming an instantaneous transition between the
    radiation and matter era (see Ref~\cite{daCunha:2021wyy} for more
    details on the approximation). For these, the Green's function are
    analytical and given by equations~\eqref{eq:greenrad} and
    \eqref{eq:greenmat}. As expected, these approximations are very
    good on sub-Hubble scales and far from the wavenumbers associated
    with the equality radiation-matter $\keq/\calHo = \order{10^2}$.}
\label{fig:k2calPh}
\end{center}
\end{figure}

The main results of this work are the strain power spectrum and the
energy density parameter of gravitational waves, evaluated today, as a
function of the wavenumber $k$. Both spectra are represented in
figure~\ref{fig:k2calPh} as black curves and have been obtained after
$30000$ hours of CPU time (on \texttt{AMD EPYC 7742}
processors). Notice that we have plotted the rescaled strain spectrum
$k^2/(12 \calHo^2)\calPh(\etanow,\etanow,k)$, which is the quantity of
interest for local measurement of GW (see
section~\ref{sec:discuss}). As discussed at length in
Ref.~\cite{daCunha:2021wyy}, this quantity is often confused with the
energy density parameter $\OmegaGW(\etanow,\etanow,k)$ whereas both
may differ (see section~\ref{sec:propsgw}).

The maximal amplitude of the spectra today occurs around the Hubble
scales. The maximal amplitude of the strain, and of the rescaled
strain, as well as the corresponding wavenumbers, read
\begin{equation}
  \begin{aligned}
  \max\left[\calPh(\etanow,\etanow,k)\right] \simeq 21 (\GU)^2 \quad
  \textrm{at} \quad k \simeq 2.0 \calHo,\\
  \max\left[\frac{k^2}{12 \calHo^2}\calPh(\etanow,\etanow,k)\right] \simeq 19 (\GU)^2  \quad
  \textrm{at} \quad k \simeq 7.9 \calHo.
  \end{aligned}
\end{equation}
The energy density parameter peaks at similar values, namely
$\max[\OmegaGW(\etanow,\etanow,k)]\simeq 8.9 (\GU)^2$ at
$k=6.7\calHo$.

In figure~\ref{fig:k2calPh}, we have also represented as dashed, and
dot-dashed, curves some semi-analytical approximations derived in
Ref.~\cite{daCunha:2021wyy}. These ones are obtained by assuming an
instantaneous transition between the radiation and matter eras such
that the Green's functions are all analytical and given by
equations~\eqref{eq:greenrad} and \eqref{eq:greenmat}. The string
correlators $\calU^{\urad}$ and $\calU^{\umat}$ are still the ones
derived from simulations while the transition era is simply
ignored. Let us notice that, as discussed in section~\ref{sec:green},
because the radiation era Green's function is inaccurate during the
matter era, the radiation-era contribution is actually derived by
integrating up to equality, and then, freely propagating the modes in
the matter era up to today. Compared to the exact solutions (black
solid curves), we see that, provided the wavenumbers are sub-Hubbles,
and not entering the Hubble radius close to equality, these
approximations are quite good while being computationally
faster. Moreover, because their sum is very close to the exact
spectrum, this confirms that GW from the transition era remain a
subdominant contribution to the overall spectra today. Only some
deviations are visible in $\OmegaGW$ in the range
$k/\calHo\in[50,500]$ (see below).

Focusing on the top panel of figure~\ref{fig:k2calPh}, after the broad
peak, let us notice the presence of a plateau at large wavenumbers,
with high frequency oscillations superimposed (black curve). Such a
plateau is indeed expected for all scaling defects evolving within the
radiation era~\cite{Figueroa:2012kw}. However, as can be seen in this
plot, the plateau is mainly sourced by the long cosmic strings
evolving within the matter era (red curve), the radiation era
contribution being significantly smaller. This matter-era plateau is
peculiar to long strings, these ones acting as singular sources of GW
according to the classification of Ref.~\cite{daCunha:2021wyy}. The
fact that their contribution is higher than the strings evolving in
the radiation era is non-trivial. For one part, it comes from the slow
decay of $\calU^{\umat}$ at large wavenumber, this is the reason why the
matter contribution is flat at high frequencies. For another part, it
comes from the higher values, taken over a wider domain, by the
correlator $\calU^{\umat}$ compared to $\calU^{\urad}$ (see
figures~\ref{fig:Urad} and \ref{fig:Umat}).

In the bottom panel of figure~\ref{fig:k2calPh}, we also see a plateau
for the energy density parameter $\OmegaGW(\etanow,\etanow,k)$, but
this one is of amplitude a few orders of magnitude smaller than the
one of the rescaled strain. Moreover, both the matter and radiation
era contributions are about the same at high wavenumber. This is in
accordance with the analytic calculations made for singular sources in
Ref.~\cite{daCunha:2021wyy} where, for a perfectly coherent correlator
scaling as $\calU^{\umat}\propto (x x')^{-1/2}$, this ratio has been
estimated as $(\etaeq/\etanow)^2$ (see figure~5 of that reference). More
intuitively, these results mean that the stochastic strain generated
by long strings is dominated by the strings close to us, the ones in
the matter era. But when considering the energy density parameter,
averaged over all space, redshifted GW emitted by the most numerous
strings of the radiation era weight as much as the close ones. From
our numerical results, we find the plateaus to have an amplitude given
by
\begin{equation}
\begin{aligned}
  \max\left[\frac{k^2}{12 \calHo^2}\calPh(\etanow,\etanow,k)\right]
&\simeq 9 (\GU)^2 \quad &\textrm{for} \quad k \gg \keq,\\
\max[\OmegaGW(\etanow,\etanow,k)] &\simeq 0.015 (\GU)^2 \quad &\textrm{for} \quad k \gg \keq,
\end{aligned}
\end{equation}
where we have introduced $\keq \equiv 1/\etaenereq \simeq 40\calHo
\simeq 0.009\,\Mpc^{-1}$. Let us notice, in the bottom panel, the
visible deviation of the exact computation (black curve) with respect
to both the matter era and radiation era semi-analytical calculations
in the range $k\in[\keq,10\keq]$. In this domain, GW from non-scaling
strings in the transition era dominate. Around $k/\calHo \simeq
10^{3}$, the exact spectrum matches again the semi-analytical
ones. Let us notice that none of the transition era effects are
visible in the strain spectrum. Indeed, at all length scales, the
strain spectrum remains dominated by the matter-era strings
contribution.

Another remark concerns the amplitude of the oscillations at high
frequencies which, in these plots, is only a few percent of the total
amplitude. As discussed in Ref.~\cite{daCunha:2021wyy}, the actual
amplitude of these oscillations depends on how fast the correlator
decays in the $\ln v$ direction, but the simulations do not allow us
to determine this very accurately at small scales. In particular, by
design, our RBF decomposition smoothes the small structures and it is
very well possible that the actual oscillation amplitude at large
wavenumbers $k/\calHo>10^3$ growths. As analytically shown in
Ref.~\cite{daCunha:2021wyy}, for the perfectly coherent correlator
$\calU^{\umat}\propto (x x')^{-1/2}$, the oscillation amplitude would
be maximal.

In summary, using the terminology of Ref.~\cite{daCunha:2021wyy}, the
spectra plotted in figure~\ref{fig:k2calPh} behave as ``constant
sources'' on super-Hubble scales (i.e., grows as $k^3$), then, for
modes of wavelength around a tenth of the Hubble radius, they transit
into a broad peak driven by the shape of the unequal time
correlator. These length scales are typical of the mean inter-string
distance in the network, and, in this region, oscillations are
washed-out by the non-vanishing off-diagonal parts of
$\calU(x,x')$. Then, around $\keq$, there is a regime associated with
non-scaling defects, apparent only in $\OmegaGW$. Finally, at large
wavenumbers, i.e., small length scales, the correlator becomes sharp
and an oscillatory pattern around a plateau appears again
(``singular'' sources). 

\subsection{Discussion}
\label{sec:discuss}
From the observational side, the GW spectra generated by long cosmic
strings are of small amplitude and only the oscillatory plateau is of
relevance for today measurements made with pulsar timing arrays and
laser interferometers ($\Ho \simeq 2.2 \times 10^{-18}\,\Hz$). From
Ref.~\cite{KAGRA:2021kbb}, the reported two-sigma upper limit
$\Omega(f=50\,\Hz) < 5.8 \times 10^{-9}$ can be converted, for long
strings, into the upper bound $\GU < 2.5 \times 10^{-5}$. In the
nanohertz range, using the strain amplitude reported in
Ref.~\cite{Antoniadis:2022pcn}, we find $k^2/(12\calHo^2)\calPh(k)
\lesssim 2.0 \times 10^{-8}$ and $\GU < 4.7 \times 10^{-5}$. These
figures are not yet competitive with the CMB
measurements~\cite{Ade:2013xla, Lazanu:2014eya}, which give $\GU <
\order{10^{-7}}$. However, let us stress again that these GW bounds
apply to long strings that would exist only in the matter era, and
these ones are essentially unconstrained by CMB
physics~\cite{Ringeval:2015ywa}. For the future GW observatories,
using the forecast $\Omega(f=10^{-3}\,\Hz) \simeq 8 \times 10^{-13}$
derived for the LISA satellites in Ref.~\cite{Boileau:2021sni}, one
would get down to a slightly better limit than CMB, namely $\GU =
10^{-7}$.

Let us clarify that, in spite of the notation, the above quantity
referred to as $\Omega(f)$ is not spatially averaged and does not
measure the cosmological energy density of GW. Rather, it is a
rescaling of the spectral density $S_h(f)$ and a measure of the local
strain power: $\Omega(f) = 4\pi^2f^3 /(3\Ho^2) S_h(f)$. It can therefore
be very different than our $\OmegaGW(\etanow,\etanow,k)$. Defining the
angular frequency $\omega = 2 \pi f$, from
equation~\eqref{eq:straincorr} and \eqref{eq:calPdef}, one has the
correspondence
\begin{equation}
  \calPh\left(\dfrac{k}{a_0} = \omega\right) \sim \dfrac{2
    \omega}{\pi} S_h(\omega),
\end{equation}
which implies that
\begin{equation}
  \Omega\left(f = \dfrac{\omega}{2\pi}\right) \sim \dfrac{k^2}{12
    \calHo^2} \calPh(k).
\end{equation}
In these expressions, the symbol ``$\sim$'' signals that we blindly
trade $k/a_0$ for $\omega$, which is an assumption. As shown in
Ref.~\cite{daCunha:2021wyy}, any given wavenumber $k/a_0$ actually
sources four angular frequencies $\omega = \pm k/a_0$ and $\omega =
\pm 2k/a_0$. The numbers quoted above assume that these angular
frequencies are all of similar amplitude.

\section{Conclusion}
\label{sec:conclusion}

Using Nambu-Goto numerical simulations of cosmic strings evolving in a
Friedmann-Lema\^{\i}tre universe, we have derived the expected
stochastic gravitational wave background (SGWB), today, generated by
the long strings all over the cosmic history. Our main results are the
strain and energy density power spectra represented in
figure~\ref{fig:k2calPh}. Although of relatively small amplitude,
typically $(k^2/\calHo^2) \calPh \simeq 100 (\GU)^2$, this signature
constitutes an absolute lower limit for most cosmic string
models. Moreover, we have found that, at high wavenumbers, the
gravitational waves emitted by strings in the matter era contribute
more to the SGWB than the ones coming from the radiation era. This is
particularly relevant for models in which strings would be reaching
scaling only within the matter era~\cite{Yokoyama:1988zza,
  Kamada:2014qta}.

For deriving our result, we have solved and propagated gravitational
waves of cosmological origin using a Green's function approach. Such
a method allows us to keep track of the time-dependence of
gravitational waves, at all times, and that is why we are able to
compute the fine oscillatory patterns in the strain and energy density
power spectra. The accuracy of the method has been discussed in
section~\ref{sec:green}, in which we have shown that using the
radiation era Green's function is incorrect as soon as an observable
mode spends some time in the super-Hubble regime. On the contrary, the
matter era Green's function, given by equation~\eqref{eq:greenmat},
has been shown to be an acceptable approximation to the $\Lambda$CDM one
at all times. Let us also notice that the treatment of the transition
era is not specific to cosmic strings. As can be seen in
equation~\eqref{eq:Imu}, the only input needed is the correlation
function $\calT(x/k,x'/k,k)$ associated with the anisotropic stress tensor. As
such, the Green's function method presented in this work is readily
applicable to any cosmological sources of gravitational waves.

Finally, this work could be extended to the $n$-point functions of the
gravitational waves, and, in particular, to the bispectrum. Cosmic
strings are genuinely non-Gaussian sources of cosmological
perturbations~\cite{Hindmarsh:2009qk, Hindmarsh:2009es} and it could
be interesting to explore the tensor side of it. The Green's function
method is well suited for this problem as the determination of the
bispectrum would involve a triple integral version of
equation~\eqref{eq:Imu}. Such a higher-dimensional problem would
certainly require to tune our radial basis functions to be separable,
in a way which has already been explored for the search of
non-Gaussianities elsewhere~\cite{Fergusson:2009nv, Fergusson:2010dm,
  Regan:2015cfa, Shiraishi:2019exr}.

\section*{Acknowledgements}
This work is supported by the ``Fonds de la Recherche Scientifique -
FNRS'' under Grant $\mathrm{N^{\circ}T}.0198.19$ as well as by the
Wallonia-Brussels Federation Grant ARC
$\mathrm{N^{\circ}}19/24-103$. Computing support has been provided
by the CURL cosmo development cluster and the Center for High Performance
Computing and Mass Storate (CISM) at UCLouvain.

\bibliographystyle{JHEP}

\bibliography{gwstg}

\end{document}